\documentclass[showpacs,preprintnumbers,amsmath,amssymb]{revtex4}
\usepackage{graphicx}
\begin{document}

\date{\today}
\vspace{2.7in}

\title{Bose-Einstein condensation of  polaritons in graphene in a high magnetic field}
\author{Oleg L. Berman$^{1}$, Roman Ya. Kezerashvili$^{1,2}$, and Yurii E.
Lozovik$^{3}$} \affiliation{\mbox{$^{1}$Physics Department, New York
City College of Technology, The
City University of New York,} \\
Brooklyn, NY 11201, USA \\
\mbox{$^{2}$The Graduate School and University Center, The
City University of New York,} \\
New York, NY 10016, USA \\
\mbox{$^{3}$Institute of Spectroscopy, Russian Academy of
Sciences,} \\
142190 Troitsk, Moscow Region, Russia}

\begin{abstract}

The Bose-Einstein condensation (BEC) of magnetoexcitonic polaritons
in a graphene layer embedded in a optical microcavity in a high
magnetic field $B$ is predicted. The essential property of this system (in contrast, e.g., to a quantum well embedded in a cavity) is stronger influence of magnetic field and weaker influence of disorder.   A two-dimensional (2D)  magnetoexcitonic polaritons gas is considered in a planar harmonic electric field potential applied to excitons or a parabolic shape of the optical cavity causing the trapping of microcavity photons. It is
shown that the effective polariton mass $M_{\rm eff}$ increases with
 magnetic field as $B^{1/2}$. The BEC
critical temperature $T_{c}^{(0)}$ decreases as $B^{-1/4}$ and
increases with  the spring constant of the parabolic
trap. The Rabi splitting related to the creation of a magnetoexciton in a high magnetic field in graphene  is obtained.
\vspace{0.1cm}



\end{abstract}

\maketitle {}



In the past decade,  Bose coherent effects of 2D excitonic
polaritons in a quantum well embedded in a semiconductor microcavity have been the subject of theoretical and
experimental studies  \cite{book}.  To obtain polaritons, two mirrors placed opposite each other form a microcavity, and quantum wells are embedded within the cavity at the antinodes of the confined
optical mode. The resonant exciton-photon interaction results in the Rabi splitting of the excitation spectrum.  Two polariton branches appear in the spectrum due to the resonant exciton-photon coupling. The lower polariton (LP) branch of the spectrum has a minimum at zero momentum.
  Recently,  the  polaritons in a harmonic potential trap have  been studied experimentally in a GaAs/AlAs quantum well embedded in a GaAs/AlGaAs microcavity \cite{science}. In this trap,
   the exciton energy is shifted using a stress-induced
band-gap. In this system, evidence for the BEC of polaritons in a quantum well
has been observed  \cite{science}. The theory of the
BEC and superfluidity of exciton polaritons in a quantum well (QW)
without magnetic field in a parabolic trap has been developed in
Ref.~[\onlinecite{Berman_L_S}].

A novel type of 2D electron system was experimentally obtained in  graphene,
 which is a 2D honeycomb lattice of the carbon atoms that form the basic planar structure in graphite
 \cite{Novoselov1,Zhang1}.
 Graphene is a gapless
semiconductor with massless electrons and holes which have been
described as Dirac-fermions.
The energy spectrum and the wavefunctions of electron-hole pairs in a magnetic field, in graphene have been calculated in Ref.~[\onlinecite{Iyengar,Koinov}].
In high magnetic field electron-hole pairs in graphene form two-dimensional magnetoexcitons.  The BEC and superfluidity of spatially-indirect magnetoexcitons with
spatially separated electrons and holes have been studied in
graphene bilayer \cite{Berman_L_G} and graphene superlattice
\cite{Berman_K_L}. The electron-hole pair condensation in the
graphene-based bilayers have been studied in~[\onlinecite{Sokolik,MacDonald2,Efetov,Joglekar}].  The  similar effects in the system of  spatially-indirect excitons (or electron-hole pairs) in coupled quantum wells (CQWs), with and without  a magnetic field  were studied in Refs.~[\onlinecite{Lozovik,Shevchenko,Berman,Poushnov,Berman_Tsvetus,Ovchinnikov}].
 The experimental and theoretical interest to  study these systems is  particularly due to the possibility of the BEC
and superfluidity of indirect excitons, which
can manifest  in the CQW as persistent electrical currents
in each well and also through coherent optical properties and
Josephson phenomena~\cite{Lozovik,Shevchenko,Berman,Poushnov,Berman_Tsvetus,Ovchinnikov}. The great experimental success was achieved now in this field~\cite{Snoke,Butov,Timofeev,Eisenstein}. Besides, the essential experimental progress was achieved in experimental studies of exciton polaritons in the system of a QW embedded in optical microcavity \cite{Dang,Yamamoto,Baumberg}. However, while the exciton polaritons have been studied in a QW, the formation of the polaritons in graphene in high magnetic field have not yet been considered. Moreover, the polaritons formed as superposition of magnetoexcitons and cavity photons in magnetic field have not yet been studied. It is interesting to study a 2D system of polaritons in graphene embedded in a microcavity from the point of view of the existence of the BEC within it. The BEC of polaritons in high magnetic field in graphene embedded in a microcavity is perspective, since the random field in graphene is weaker than in a QW, particularly, because in a QW the random field is generated due to the fluctuations of the width of a QW.  Let us mention that if the interaction of bosons with the random field is stronger, the BEC critical temperature is lower \cite{BLSC}. Moreover, the BEC of polaritons in graphene embedded in a microcavity can exist at much lower magnetic field than in a QW, because the distance between electron Landau levels in graphene is much higher than in a QW at the same magnetic field, and, therefore, the lower magnetic field is required in graphene than in a QW to neglect the electron transitions between the Landau levels. The purpose of this Letter is to point out the existence of the BEC of the magnetoexciton polaritons in a graphene layer embedded in a semiconductor microcavity in a strong magnetic field and to discuss the conditions of its realization.

When an undoped electron system in graphene in a magnetic field without an external electric field is in the ground state, half of the zeroth Landau level is filled with electrons, all Landau levels above the zeroth one are  empty, and all levels below  the zeroth one are filled with electrons. We suggest using the gate voltage to control the chemical potential in graphene by two ways: to shift it above the zeroth level so that it is between the zeroth and first Landau levels (the first case) or to shift the chemical potential below the zeroth level so that it is between the  first negative and zeroth Landau levels (the second case). In both cases, all Landau levels below the chemical potential are completely filled and all Landau levels above the chemical potential are completely empty. Based on the selection rules for optical transitions between the Landau levels in  single-layer graphene~\cite{Gusynin3}, in the first case, there are allowed transitions between the zeroth and the first Landau levels, while in the second case there are allowed transitions between the first negative  and zeroth Landau levels.
  Correspondingly, we consider magnetoexcitons formed in graphene by the electron on the first Landau level and the hole on the zeroth Landau level (the first case) or the electron on the zeroth Landau level and the hole on the Landau level $-1$ (the second case). Note that by appropriate gate potential we can also use  any other neighboring Landau levels $n$ and $n+1$.

 For the relatively high dielectric constant of the microcavity,   $\epsilon \gg e^{2}/(\hbar v_{F}) \approx 2$ ($v_{F} = \sqrt{3}at/(2\hbar)$  is the Fermi velocity of electrons in graphene, where
$a =2.566 \ \mathrm{\AA}$  is a lattice constant and $t \approx 2.71 \
\mathrm{eV}$ is the overlap integral between the nearest carbon
atoms \ \cite{Lukose}) the magnetoexciton energy in graphene can be
calculated by applying perturbation theory with respect to the strength of the
Coulomb electron-hole attraction analogously as it was done in [\onlinecite{Lerner}] for 2D quantum wells in a high magnetic field with non-zero electron $m_{e}$
and hole $m_{h}$ masses. This approach
allows us to obtain the spectrum
  of an isolated magnetoexciton with the electron on the Landau level $1$ and the hole on the Landau level $0$ in a single graphene layer, and it will be exactly the same as for the magnetoexciton with the electron on the Landau level $0$ and the hole on the Landau level $-1$.
The characteristic Coulomb electron-hole attraction for the single
graphene layer is $e^{2}/(\epsilon r_{B})$,  where $\epsilon $ is the dielectric constant of the
environment around graphene, $r_{B} = \sqrt{c\hbar/(eB)}$ denotes
the magnetic length of the magnetoexciton in the magnetic field $B$, and $c$ is the speed of light. The
energy difference between the first and zeroth Landau levels in
graphene is  $\hbar v_{F}/r_{B}$.    For graphene, the perturbative approach with respect to the strength of the Coulomb electron-hole attraction  is valid when $e^{2}/(\epsilon r_{B}) \ll \hbar
v_{F}/r_{B}$ \cite{Lerner}.
  This condition can be fulfilled at all magnetic
fields $B$ if the dielectric constant of the surrounding media satisfies the condition $
e^{2}/(\epsilon \hbar v_{F}) \ll 1$. Therefore, we claim that the
energy difference between the first and zeroth Landau levels is
always greater than the characteristic Coulomb attraction between the
electron and the hole in the single graphene layer at any $B$ if $\epsilon \gg e^{2}/(\hbar v_{F}) \approx 2$. Thus, applying perturbation theory with respect to weak Coulomb electron-hole attraction in graphene embedded in the $\mathrm{GaAs}$ microcavity ($\epsilon = 12.9$) is more accurate than for graphene embedded in the $\mathrm{SiO}_{2}$ microcavity ($\epsilon = 4.5$).  However, the magnetoexcitons in graphene exist in high magnetic field. Therefore, we restrict ourselves by consideration of high magnetic fields.

Polaritons are linear superpositions of excitons and photons. In
high magnetic fields, when magnetoexcitons may exist, the polaritons
become linear superpositions of magnetoexcitons and photons. Let
us define the superpositions of magnetoexcitons and photons as
magnetopolaritons.  It is obvious that magnetopolaritons in graphene
are two-dimensional, since graphene is a two-dimensional structure.
The Hamiltonian of magnetopolaritons in the strong magnetic field is
given by
\begin{eqnarray}
\label{Ham_tot_pol} \hat{H}_{tot} = \hat{H}_{mex} + \hat{H}_{ph} +
\hat{H}_{mex-ph} \ ,
\end{eqnarray}
where $\hat{H}_{mex}$ is a magnetoexcitonic Hamiltonian,
 $\hat{H}_{ph}$ is a photonic Hamiltonian, and $\hat{H}_{exc-ph}$ is the
Hamiltonian of magnetoexciton-photon interaction.

Let us analyze each term of the Hamiltonian for magnetopolaritons
(\ref{Ham_tot_pol}). The effective Hamiltonian and the energy dispersion for magnetoexcitons  in graphene layers in a high magnetic field derived in Ref.~[\onlinecite{Berman_K_L}]  are given by
\begin{eqnarray}
\label{sp_exc} \hat H_{mex} = \sum_{{\bf P}}^{}\varepsilon_{mex}(P)
\hat{b}_{{\bf P}}^{\dagger}\hat{b}_{{\bf P}}^{} \ , \ \ \ \ \ \ \ \
\varepsilon_{mex}(P) = E_{band} -
\mathcal{E}_{B}^{(b)} + \varepsilon _{0}(P) \ ,
\end{eqnarray}
 where $\hat{b}_{{\bf P}}^{\dagger}$ and $\hat{b}_{{\bf P}}$ are magnetoexcitonic creation
and annihilation operators obeying the Bose commutation relations.
In Eq.~(\ref{sp_exc}), $E_{band}= E_{1,0}^{(0)} = \sqrt{2}\hbar v_{F}/r_{B}$ is the band
gap  energy, which is the difference between the Landau levels $1$ and $0$ in graphene. $\mathcal{E}_{B}^{(b)}$ is the binding energy of a 2D
magnetoexciton with the electron in the Landau level $1$ and the hole on
the Landau level $0$ in a single graphene layer, and $\varepsilon
_{0}(P) = P^2/(2m_{B})$, where $m_{B}$ is the effective magnetic
mass of a 2D magnetoexciton with the electron on the Landau level
$1$ and hole on the Landau level $0$ in a single graphene layer. The binding energy $\mathcal{E}_{B}^{(b)}$ and effective magnetic mass $m_{B}$ of a magnetoexciton in graphene obtained using the first order perturbation respect to the electron-hole Coulomb attraction similarly to the case of a single quantum well \cite{Lerner} are given by
\begin{eqnarray}\label{emsa}
\mathcal{E}_{B}^{(b)} =   \sqrt{\frac{\pi}{2}}  \frac{e^{2}}{\epsilon r_{B}} ,
\hspace{3cm} m_{B} = \frac{2^{7/2}\epsilon \hbar^{2}}{\sqrt{\pi}e^{2}r_{B}} \ .
\end{eqnarray}

It can be shown that the interaction between two direct 2D
magnetoexcitons in graphene with the electron on the Landau level $1$ and the hole on the Landau level $0$ can be
neglected in a strong magnetic field, in analogy to what is described in Ref.~[\onlinecite{Lerner}] for 2D magnetoexcitons in a quantum well. The dipole moment
of each exciton in a magnetic field is $\mathbf{d}_{1,2} =
e\mathbf{\rho}_{0} =r_{B}^{2}\left[\mathbf{B}\times
\mathbf{P}_{1,2}\right]/B$ \cite{Lerner}, where
 $\mathbf{P}_{1}$ and $\mathbf{P}_{2}$ are the magnetic momenta of each exciton and $P_{1}, P_{2}\ll 1/r_{B}$. The magnetoexcitons are located
 at a distance $R \gg r_{B}$ from each other. The corresponding contribution to the energy of
  their dipole-dipole interaction is
   $\sim \mathcal{E}_{B}^{(b)} \left(r_{B}/R\right)^{3}P_{1}P_{2}r_{B}^{2}/\epsilon \sim \left(r_{B}/R\right)^{3}P_{1}P_{2}/(\epsilon M_{0}) \ll  e^{2}r_{B}^{2}/(\epsilon R^{3})$.
   Inputting the radius of the magnetoexciton in graphene   $r_{0,1} \sim r_{B}$
   \cite{Berman_L_G}, we obtain that
    the van der Waals attraction of
    the exciton at zero momenta is proportional to $\sim \left(r_{0,1}/R\right)^{6} \sim \left(r_{B}/R\right)^{6}$.
    Therefore, in the limit of a strong magnetic field for a dilute system
     $r_{B}\ll R$, both the dipole-dipole interaction and the van der Waals attraction vanish,
     and the 2D magnetoexcitons in graphene form an ideal Bose gas analogously to the 2D magnetoexcitons in a quantum well given in Ref.~[\onlinecite{Lerner}].
    Thus, the Hamiltonian (\ref{Ham_tot_pol}) does not include
     the term corresponding to the interaction between two direct
     magnetoexcitons in a single graphene layer.  So in high magnetic field  there is the BEC of the ideal magnetoexcitonic gas in
     graphene.

The Hamiltonian and the energy spectrum of non-interacting photons in a semiconductor
microcavity are given by  \cite{Pau}:
\begin{eqnarray}
\label{sp_phot} \hat H_{ph} = \sum_{{\bf P}}  \varepsilon _{ph}(P)
\hat{a}_{{\bf P}}^{\dagger}\hat{a}_{{\bf P}}^{} \ , \ \ \ \ \ \ \
\varepsilon _{ph}(P) = (c/n)\sqrt{P^{2} +
\hbar^{2}\pi^{2}L_{C}^{-2}} \ ,
\end{eqnarray}
 where $\hat{a}_{{\bf P}}^{\dagger}$ and $\hat{a}_{{\bf P}}$ are
photonic creation and annihilation Bose operators.
In Eq.~(\ref{sp_phot}), $L_{C}$ is the length of the cavity, $n =
\sqrt{\epsilon_{C}}$ is the effective refractive index and
$\epsilon_{C}$ is the dielectric constant of the cavity. We assume that the
length of the microcavity has the following form:
\begin{eqnarray}
\label{lc} L_{C}(B) = \frac{\hbar\pi c}{n \left(E_{band} -
\mathcal{E}_{B}^{(b)}\right)} \ ,
\end{eqnarray}
corresponding to the resonance of the photonic and magnetoexcitonic branches at $P = 0$, i.e.  $\varepsilon_{mex}(0) = \varepsilon_{ph}(0)$. As it follows from the energy spectra in~(\ref{sp_exc}) and~(\ref{sp_phot}), and Eqs.~(\ref{emsa}) and~(\ref{lc}),
 the length of the microcavity, corresponding to a
magnetoexciton-photon resonance, decreases with the increment of
the magnetic field as $B^{-1/2}$. The resonance between
magnetoexcitons and cavity photonic modes can be achieved either by
controlling the spectrum  of magnetoexcitons $\varepsilon_{ex}(P)$
by changing magnetic field $B$ or by choosing  the appropriate length  of the
microcavity $L_{C}$.

 The Hamiltonian of the magnetoexciton-photon coupling has the form (see Refs.~[\onlinecite{Agranovich1,Ciuti,Agranovich2}]):
\begin{eqnarray}
\label{Ham_exph} \hat{H}_{mex-ph} = {\hbar \Omega_{R}}\sum_{{\bf P}}
 \hat{a}_{{\bf P}}^{\dagger}\hat{b}_{{\bf P}}^{} + h.c. \ ,
\end{eqnarray}
\begin{eqnarray}
\label{bog_tr} \hat{b}_{\mathbf{P}} = X_{P}\hat{p}_{\mathbf{P}} -
C_{P}\hat{u}_{\mathbf{P}}, \hspace{3cm}
 \hat{a}_{\mathbf{P}} = C_{P}\hat{p}_{\mathbf{P}} +
X_{P}\hat{u}_{\mathbf{P}} \ ,
\end{eqnarray}
where the magnetoexciton-photon coupling energy represented by the
Rabi constant $\hbar \Omega_{R}$ is obtained below,   $\hat{p}_{\mathbf{P}}$ and $\hat{u}_{\mathbf{P}}$ are lower and upper magnetopolariton Bose
operators, respectively, j  $X_{P}$ and $C_{P}$ are coefficients of the unitary Bogoliubov transformation~\cite{Agranovich1,Agranovich2}, and the energy spectra of the lower/upper magnetopolaritons are
\begin{eqnarray}
\label{eps0} \varepsilon_{LP/UP}(P) &=& \frac{\varepsilon _{ph}(P)
+\varepsilon _{mex}(P)}{2}  \nonumber \\ &\mp&
\frac{1}{2}\sqrt{(\varepsilon_{ph}(P) - \varepsilon _{mex}(P))^{2} +
4|\hbar\Omega_{R}|^{2}} \ .
\end{eqnarray}
Eq.~(\ref{eps0}) implies a splitting of $2 \hbar \Omega_R$ between the upper and lower
states of polaritons at $P=0$, which is known as the
Rabi splitting.

 Substituting~(\ref{bog_tr}) into Hamiltonians~(\ref{sp_exc}),~(\ref{sp_phot}) and~(\ref{Ham_exph}) and diagonalizing the Hamiltonian~(\ref{Ham_tot_pol}), finally we obtain
\begin{eqnarray}
\label{h0} \hat{H}_{tot} =
\sum_{\mathbf{P}}\varepsilon_{LP}(P)\hat{p}_{\mathbf{P}}^{\dagger}\hat{p}_{\mathbf{P}}
+\sum_{\mathbf{P}}\varepsilon_{UP}(P)\hat{u}_{\mathbf{P}}^{\dagger}\hat{u}_{\mathbf{P}},
\end{eqnarray}
where $\hat{p}_{\mathbf{P}}^{\dagger}$, $\hat{p}_{\mathbf{P}}$,
$\hat{u}_{\mathbf{P}}^{\dagger}$,  $\hat{u}_{\mathbf{P}}$ are the
Bose creation and annihilation operators for the lower and upper
magnetopolaritons, respectively. The Hamiltonian
Eq.~(\ref{h0}) describes  magnetopolaritons in a single
graphene layer in a high magnetic field. Our particular interest is
the lower energy magnetopolaritons which produce the BEC. The lower
palaritons have the lowest energy within a single graphene layer.

Similarly to the  Bose atoms in a
trap in the case of a slowly varying
external potential \cite{Pitaevskii},   we can make the quasiclassical approximation,
 assuming that the effective magnetoexciton mass does not depend on a characteristic size $l$  of the trap and it is a constant within the trap.
  This quasiclassical approximation is valid if $P \gg \hbar/l$. The harmonic trap is formed by the two-dimensional planar potential in the plane of graphene.
  The potential trap can be produced in two
different ways. In case 1, the potential trap can be produced by
applying an external inhomogeneous electric field. The spatial
dependence of the external field potential $V(r)$ is caused by
shifting of  magnetoexciton energy by applying an external
inhomogeneous electric field. The photonic states in the cavity are
assumed to be unaffected by this electric field. In this case the
band energy $E_{band}$ is replaced by $E_{band}(r) = E_{band}(0)+
V(r)$. Near the minimum of the magnetoexciton energy, $V(r)$ can be
approximated by the planar harmonic potential $\gamma r^{2}/2 $,
where $\gamma$ is the spring constant. Note that a high magnetic field
does not change the
 trapping potential in the effective Hamiltonian \cite{Berman_L_S_C}. In
case 2, the trapping of magnetopolaritons is caused by the
inhomogeneous shape of the cavity when the length of the cavity is determined by Eq.~(\ref{lc}) with the term $\gamma r^{2}/2$ added to $E_{band} - \mathcal{E}_{B}^{(b)}$,
where $r$ is the distance between the photon
 and the center of the trap. In case 2, the $\gamma$  is the curvature characterizing the shape of the cavity.  In case 1, for the slowly changing confining potential $V(r) =
\gamma  r^{2}/2$, the magnetoexciton
spectrum is    given in the effective mass approximation by Eq.~(\ref{sp_exc}) provided we added  the term $\gamma r^{2}/2$ added to $E_{band} - \mathcal{E}_{B}^{(b)}$ in the r.h.s.,
where $r$ is now the distance between the center of mass of
the   magnetoexciton and the center of the trap.
The Hamiltonian and the energy spectrum of the photons in this case are shown by
Eq.~(\ref{sp_phot}), and the length of the microcavity is given by
Eq.~(\ref{lc}).

In case 2, for  the photonic spectrum in the effective mass approximation is given by  substituting the slowly changing shape of the length of cavity depending on the term  $\gamma r^{2}/2$ into Eq.~(\ref{sp_phot}) representing the spectrum of the cavity photons.
This quasiclassical approximation is valid if $P \gg \hbar /l$,
where  $l = \left(\hbar/(m_{B}\omega_{0})\right)^{1/2}$ is the size
of the magnetoexciton cloud in an ideal magnetoexciton gas and
$\omega_{0} = \sqrt{\gamma/m_{B}}$. The Hamiltonian and energy spectrum of
magnetoexcitons in this case are given by~(\ref{sp_exc}).

At small momenta $\alpha \equiv
1/2 (m_{B}^{-1} + (c/n)L_{C}/\hbar\pi)P^{2}/|\hbar \Omega_{R}| \ll
1$ ( $L_{C} = \hbar\pi c/n \left(E_{band} -
\mathcal{E}_{B}^{(b)}\right)^{-1}$) and weak confinement $\beta
\equiv \gamma r^{2}/|\hbar \Omega_{R}| \ll 1$, the single-particle
lower magnetopolariton spectrum obtained through the substitution of
Eq.~(\ref{sp_exc}) into Eq.~(\ref{eps0}), in linear order with
respect to the small parameters $\alpha$ and $\beta$, is
\begin{eqnarray}
\label{eps00} \varepsilon_{0}(P) \approx \frac{c}{n} \hbar \pi
L_{C}^{-1} - |\hbar \Omega_{R}| +  \frac{\gamma}{4} r ^{2} +
\frac{1}{4} \left(m_{B}^{-1} + \frac{c
L_{C}(B)}{n\hbar\pi}\right)P^{2} \ .
\end{eqnarray}
Let us emphasize that the spectrum of non-interacting
magnetopolaritons $\varepsilon_{0}(P)$  at small momenta and weak
confinement is given by Eq.~(\ref{eps00}) for  both physical
realizations of confinement: case 1 and case 2.

If we measure the energy relative to the $P=0$ lower magnetopolariton
energy $(c/n) \hbar \pi L_{C}^{-1}  - |\hbar \Omega_{R}|$, we obtain
the resulting  effective Hamiltonian for trapped magnetopolaritons
in graphene in a magnetic field. At small momenta $\alpha \ll 1$
($L_{C} = \hbar\pi c/n \left(E_{band} -
\mathcal{E}_{B}^{(b)}\right)^{-1}$) and weak confinement $\beta
 \ll 1$, this effective Hamiltonian is
\begin{eqnarray}
\label{Ham_eff} \hat H_{\rm eff}  =
\sum_{\mathbf{P}}\left(\frac{P^{2}}{2M_{\rm eff}(B)} + \frac{1}{2}
V(r) \right)\hat{p}_{\mathbf{P}}^{\dagger}\hat{p}_{\mathbf{P}} \ ,
\end{eqnarray}
 where the sum over $\mathbf{P}$  is carried out only over $P \gg \hbar/l$
  (only in this case the quasiclassical approach  used in Eq.~(\ref{sp_exc}) is valid), and
 the effective magnetic mass of a magnetopolariton is
given by
\begin{eqnarray}
\label{Meff} M_{\rm eff}(B) = 2   \left(m_{B}^{-1} + \frac{c
L_{C}(B)}{n\hbar\pi}\right)^{-1} \ .
\end{eqnarray}
 According to Eqs.~(\ref{Meff}) and~(\ref{emsa}), the effective
magnetopolariton mass $M_{\rm eff}$ increases with the increment of
the magnetic field as $B^{1/2}$.
 Let us emphasize that the resulting effective Hamiltonian for magnetopolaritons  in graphene in a magnetic field for the parabolic trap is given by
Eq.~(\ref{Ham_eff}) for both physical realizations of
confinement represented by case 1 and case 2.

Neglecting anharmonic terms for the magnetoexciton-photon coupling,  the Rabi splitting constant $\Omega_{R}$  can
be estimated quasiclassically as
\begin{eqnarray}
\label{defrabi}
\left|\hbar \Omega_{R}\right|^{2} =  \left| \left\langle  f \left|\hat{H}_{int} \right| i \right\rangle  \right|^{2} \ ,  \ \ \ E_{ph0} = \left(\frac{2 \pi \hbar \omega}{\epsilon W} \right)^{1/2} \ ,
\ \ \  \hat{H}_{int}  = -  \frac{v_{F}e}{c} \vec{\hat{\sigma}}\cdot \vec{A}  =   \frac{v_{F}e}{i \omega} \vec{\hat{\sigma}}\cdot \vec{E}_{ph0} \ ,
\end{eqnarray}
where $\vec{\hat{\sigma}} = (\hat{\sigma}_{x},\hat{\sigma}_{y})$, $\hat{\sigma}_{x}$ and $\hat{\sigma}_{y}$ are Pauli matrices, $\hat{H}_{int}$ is the Hamiltonian of the electron-photon interaction corresponding to the electron in graphene described by Dirac dispersion, $E_{ph0}$ is the electric field corresponding to a single cavity photon,
 $W$ is the volume of microcavity,   $\omega$ is the photon frequency. The initial $| i \rangle$ electron state corresponds to the completely filled  Landau level $0$ and completely empty  Landau level $1$. The final $| f \rangle$ electron state corresponds to creation of one magnetoexciton with the electron on the Landau level $1$ and the hole on the Landau level $0$.
The transition dipole moment corresponding to the process of creation of this magnetoexciton is given by $d_{12} = e r_{B}/4$.
Let us note that in Eq.~(\ref{defrabi}) the energy of photon  absorbed at the creation of the magnetoexciton is given by $\hbar \omega = \varepsilon_{1} - \varepsilon_{0}  = \sqrt{2} \hbar v_{F}/r_{B}$ (we assume that $\mathcal{E}_{B}^{(b)} \ll \varepsilon_{1} - \varepsilon_{0}$).
Substituting the photon energy  and the transition dipole moment from  into   Eq.~(\ref{defrabi}), we obtain the Rabi splitting corresponding to the creation of a magnetoexciton with the electron on the Landau level $1$ and the hole on the Landau level $0$ in graphene: $\hbar \Omega_{R} = e \left( \pi \hbar v_F r_{B}/(\sqrt{2} \epsilon W) \right)^{1/2}$.

Thus, the Rabi splitting in graphene is related to the creation of the magnetoexciton, which decreases when the magnetic field increases and is proportional to $B^{-1/4}$. Therefore, the Rabi splitting in graphene can be controlled by the external magnetic field.
It is easy to show that the Rabi  splitting related to the creation of the magnetoexciton, the electron on the Landau level $0$ and the hole on the Landau level $-1$ will be exactly the same as for the magnetoexciton with the electron on the Landau level $1$ and the hole on the Landau level $0$.
Let us mention that dipole optical transitions from the Landau level $-1$ to the  Landau level $0$, as well as from  the Landau level $0$ to the  Landau level $1$, are allowed by the selection rules for optical transitions in single-layer graphene \cite{Gusynin3}.

\begin{figure}[t] 
   \centering
  \includegraphics[width=3.4in]{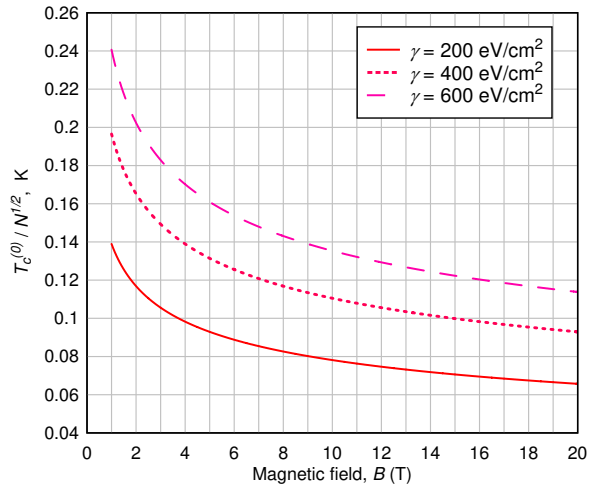}
   \caption{The ratio of the BEC critical temperature to the square root of the total number of magnetopolaritons $T_{c}^{(0)}/\sqrt{N}$ as a function of magnetic
   field $B$  at different spring constants $\gamma$. We assume the environment around graphene is $\mathrm{GaAs}$ with $\epsilon = 12.9$.}
   \label{fig_t_c}
\end{figure}

As known, the exact solution for the ground state of 2D magnetoexcitons in  high magnetic fields is the Bose condensate of non-interacting magnetoexcitons~\cite{Lerner}. This is valid for both semiconductor quantum well and graphene.
Although Bose-Einstein condensation cannot  take place in a 2D
homogeneous ideal gas at non-zero temperature, as discussed in
Ref.~[\onlinecite{Bagnato}], in a harmonic trap the BEC can occur in two
dimensions below a critical temperature $T_{c}^{0}$. In a harmonic
trap at a temperature $T$ below a critical temperature $T_{c}^{0}$ ($T
< T_{c}^{0}$), the number $N_{0}(T,B)$ of non-interacting magnetopolaritons in
the condensate  is given in Ref.~[\onlinecite{Bagnato}].
Applying the condition $N_{0}=0$, and assuming
that the magnetopolariton effective mass  is given by Eq.~(\ref{Meff}),
we obtain the BEC critical temperature $T_{c}^{(0)}$ for the ideal
gas of magnetopolaritons in a single graphene layer  in a magnetic
field:
\begin{eqnarray}
 \label{t_c}
T_{c}^{(0)} (B) = \frac{1}{k_{B}}\left(\frac{3\hbar^{2}\gamma N}{\pi
 \left(g_{s}^{(e)}g_{v}^{(e)} +
 g_{s}^{(h)}g_{v}^{(h)}\right)M_{\rm eff}(B)}
\right)^{1/2} \ ,
\end{eqnarray}
where $N$ is the total number of magnetopolaritons,
$g_{s}^{(e),(h)}$ and $g_{v}^{(e),(h)}$ are the spin and graphene
valley degeneracies for an electron and a hole, respectively,
$k_{B}$ is the Boltzmann constant.
At temperatures above $T_{c}^{(0)}$,  the BEC of magnetopolaritons in a single graphene layer does not exist.
$T_{c}^{(0)}/\sqrt{N}$ as a function of magnetic
   field $B$  and spring constant $\gamma$ is presented in
   Fig.~\ref{fig_t_c}. In our calculations, we used $g_{s}^{(e)} = g_{v}^{(e)} =
g_{s}^{(h)} = g_{v}^{(h)} = 2$.
 According to Eq.~(\ref{t_c}), the BEC critical temperature
$T_{c}^{(0)}$ decreases with the magnetic field  as
$B^{-1/4}$ and increases with  the spring constant
as $\gamma^{1/2}$. Note that we assume that the quality of the cavity is sufficiently high, so that the time of the relaxation to the Bose condensate quasiequilibrium state is smaller than the life time of the photons in the cavity.

To conclude, we have studied the ideal
gas of trapped cavity magnetopolaritons in a single graphene layer
in a high magnetic field. The resonance between magnetoexcitons and
cavity photonic  modes can be achieved either by  controlling the
spectrum  of magnetoexcitons $\varepsilon_{ex}(P)$, by changing
magnetic field $B$ or by controlling the length  of  the microcavity
$L_{C}$.  We analyzed two possible physical realizations of the
trapping potential: a harmonic electric field potential coupled to
magnetoexcitons and a parabolic shape of the semiconductor cavity
causing the trapping of microcavity photons. We conclude that both
 realizations of  confinement result in the same effective
Hamiltonian. It is shown that the effective magnetopolariton mass
$M_{\rm eff}$ increases with the  magnetic field as
$B^{1/2}$. Meanwhile, the BEC critical temperature $T_{c}^{(0)}$
decreases as $B^{-1/4}$ and increases with the spring constant
 as $\gamma^{1/2}$.  The gas of magnetopolaritons in graphene in a high
magnetic field can be treated as an ideal Bose gas since
magnetoexciton-magnetoexciton interaction vanishes in the limit of
a high magnetic field and a relatively high dielectric constant of the cavity $\epsilon \gg 2$. Let us mention that this condition for the high dielectric constant of the microcavity is valid only for graphene, and it is not valid for the quantum well.   Besides, we have obtained the Rabi splitting related to the creation of a magnetoexciton in a high magnetic field in graphene which can be controlled by the external magnetic field $B$.

\acknowledgments

 O.~L.~B., R.~Ya.~K. were supported by PSC CUNY grant 621360040, and Yu.~E.~L. by RFBR grants and programs of Russian Academy of Science.



\end{document}